# HTS YBCO Resonator Configuration with Coplanar Optimized Flux Concentrator Strongly Coupled to rf SQUID

Fatemeh Qaderi, Faezeh Shanehsazzadeh, Behnam Mazdouri, Mehdi Fardmanesh

Sharif University of Technology, Thran, Iran

*Abstract*— We developed a novel magnetic coupling module formed of a monolayer superconducting flux concentrator, which is integrated with a coplanar resonator strongly coupled to HTS rf-SQUID. Three types of resonators, including a long stripline resonator between input loop and pick-up loop of the flux concentrator, a complementary split ring resonator (CSRR), and also a spiral shape inside the input loop are explored. The resonance quality factors as well as the coupling to the SQUID of different patterns of these three types of the resonators is evaluated using Finite Element Method (FEM) simulations. Several readout methods to couple the electronic system to the resonators are tested, including inductive (coil) and capacitive (transmission line) couplings, and the optimum readout is reported for each of the resonators. Among the evaluated resonator types, a spiral shape resonator with optimal design showing the highest quality factor (5900) together with the strongest coupling to the SQUID (-0.5 dB) at resonance frequency of 836 MHz, is fabricated using 200 nm thick superconducting YBCO on a 1 mm thick crystalline LaAlO3 substrate. The flux concentrator of the module is optimized by the variation of its linewidths and also its input loop radius to obtain maximum flux transformation efficiency.

## I. Introduction

SUPERCONDUCTING Quantum Interference Devices (SQUIDs) have been of great research attention for magnetometry [1], non-destructive evaluation (NDE) [2], geophysical measurements [3], bio-magnetic measurements and in particular, magneto-cardiography [4], [5] for its noninvasive nature [6].

Fabrication of HTS rf SQUID with integrated monolithic or flip-chip flux transformer in order to enhance magnetic field sensitivity of the device is reported in [7]-[9]. The Tank circuit part of the SQUID resonator module is also reported to be integrated with the flux transformer to better satisfy the requirement $k^2Q > 1$, which affects is a determining factor for the flux noise of the device [10]. Where $k$ is the coupling coefficient between the resonator and the SQUID, and Q stands for the resonator quality factor. Superconducting resonators are more favorable because of their higher quality factor compared to that of the lumped element resonators [11] and also their capability of being integrated with the flux transformer on a single substrate to achieve strong coupling to the SQUID. The use of flux transformer (or concentrator) is essential to extend the effective area sensible by the SQUID and to lower the flux noise [12].

We propose a new monolayer configuration of a magnetic coupling module, exploiting a flip-chip resonator-flux concentrator structure. This monolayer design has the advantage of avoiding complicated fabrication process, as well as higher yield and less parasitic effects compared to multilayer structures. We explore three different types of resonators and find the optimum pattern of the readout for each, with the optimization criteria of maximum quality factor and coupling at the resonance frequency. We have to keep the resonance frequency low enough so that the electronic system can follow it, in this case, below 1 GHz. Designing such a resonator at the available dimensions has many considerations, because e.g. resonators with larger sizes can resonate at the desired frequency range, but have very little coupling to the SQUID due to the mismatch between the sizes of the resonator and the rf-SQUID used [13], [14]. We compare the behavior of long stripline resonators (with resonance frequency well below 1 GHz), complementary split ring resonators (CSRRs with different radiuses, inside or outside the input loop of transformer), and a Spiral pattern (with dimensions similar to that of the SQUID used).

## II. Flux Concentrator Optimization

The magnetic field sensitivity of the SQUID is enhanced by a superconducting flux transformer (or concentrator), through extension of the effective area sensible by the SQUID. The effective area is as follows in Eq. (1):

F. Qaderi, F. Shanehsazzadeh and M. Fardmanesh are with Superconductive Electronics Research Lab (SERL) at Sharif University of Technology, Tehran, Iran. B. Mazdouri is with Shahed University, Tehran, Iran.
(email: fatemeh.qaderi@epfl.ch; f_shanehsazzadeh@ee.sharif.edu; behnam.mazdouri@yahoo.com; fardmanesh@sharif.edu)

$$\frac{\partial B}{\partial \varphi} = \frac{1}{A_{eff}} = \frac{L_i + L_p}{A_s(L_i + L_p) + A_p \alpha \sqrt{L_s L_i}} \quad (1)$$

Where $A_s$ is the bare SQUID effective area, $L_s$ the SQUID inductance, $L_i$ the input loop inductance, $L_p$ the pick-up loop inductance, $A_p$ the pick-up area and $\alpha$ the coupling coefficient [12], [15].

We propose an optimized monolayer flux concentrator for strong coupling to the SQUID. It is similar to the well-known single-turn transformer with the input loop folded inside the pick-up loop (Figure 1). In case of matching inductances of the two loops, the flux is transformed more effectively [16], [17].

Excluding the pick-up loop from the structure, the inductance of the input loop is simulated in sweeps on its radius as well as its linewidth. Similar simulations are performed for the pick-up loop linewidth, at a fixed outer side of the square of 8 mm. The square size is not swept, because the pick-up loop should take the maximum affordable area with respect to the substrate, in order to minimize the leakage in magnetic field detectivity. The behavior of the input loop inductance versus its diameter is demonstrated in Figure 2(a), at a constant linewidth of 100 μm. Figure 2(b) shows the variations of input loop inductance according to its linewidth, decreasing inductance by increasing the linewidth. The inductance of a square-shaped pick-up loop, changing with the linewidth from 200 μm to 1.2 mm is also shown in Figure 2(c). Results are consistent with the following Eq. 2:

$$L_{loop} = \mu_r \mu_0 r (\ln(8r/w) - 2) \quad (2)$$

Where $r$ is the average radius of the loop and $w$ is its linewidth [18].

According to these sweeps results, the inductances of the input loop and pick-up loop do not meet at these dimensions. Moreover, when the SQUID is added to the structure in a flip-chip configuration, it confines the input loop flux and causes its inductance to decrease (like the effect of a ground plane on the inductance of a slab). However, to keep their inductances as close as possible and based on the fabrication considerations, the linewidth of 1 mm and 100 μm for pick-up and input loop are chosen, respectively.

In the next step, considering the SQUID effect, the overall efficiency of the transformer is evaluated by sweeping the input loop diameter, at fixed linewidths of the pick-up loop and input loop equal to 1 mm and 100 μm, respectively. The transformer of Figure 1 together with the SQUID in a flip-chip configuration is placed in a magnetic field excitation perpendicular to the transformer plane, generated by an external port (port 1). Another port (port 2) is placed instead of the SQUID's Josephson junction and the S21 parameter is considered as a criterion of flux transformation efficiency. Figure 2(d) shows the behavior of S21 versus input loop diameter changing from 1 to 4 mm (Note: The S21 values of Figure 2(d) are just comparative, the absolute values do not carry accurate information, as there is some constant leakage about 10 dB before the transformer picks up the generated flux). There are two trends acting in reverse in Figure 2(d): First, increasing the input loop diameter, due to increasing its inductance, enhances the inductance matching of the two loops. Second, smaller input loop results in higher concentration of magnetic flux at the center, i.e. where the junction is located [19]. These trends explain how an optimum diameter is found around 2.5 mm, as shown in Figure 2(d).

The field to flux transformation coefficient ($\partial B/\partial \varphi$) for this transformer when loaded by SQUID, is about 1.47 nT/$\Phi_0$, showing a drop from the bare SQUID case (9 nT/$\Phi_0$). This value is desired to be as low as possible [12].

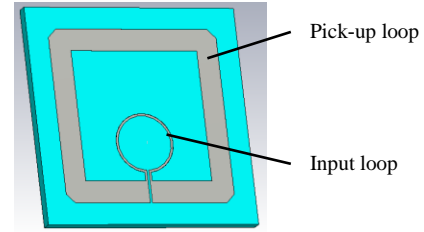

Figure 1 Flux transformer schematic

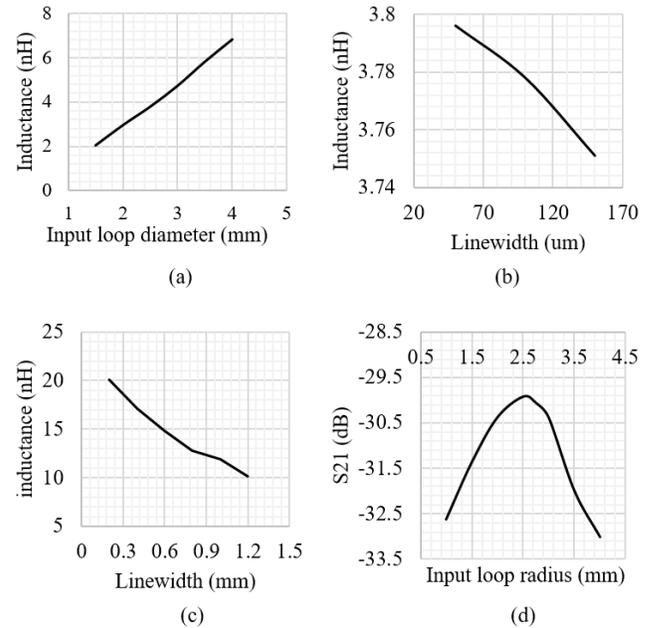

Figure 2. (a) Input loop inductance vs. its diameter, for a loop with linewidth of 100 μm, (b) Input loop inductance vs. linewidth, for a circular loop with diameter of 2.5 mm, (c) Pick-up loop inductance vs. linewidth, for a semi-square shape with average side of 8mm, (d) S21 as a measure of flux transformation efficiency, vs. input loop diameter

### III. RESONATOR PATTERN

There are different candidates for the resonator pattern: ring resonators [14], [20], or the new proposed spiral shapes and other possible arbitrary patterns as shown in Fig. 3. Each of these candidates needs a special configuration when being integrated with the flux transformer. The important consideration of the design is that the resonance frequency should not go higher than the maximum affordable frequency of the readout electronic system used, which is around 1 GHz. This may require the resonator size to go larger than the available area of our typical LAO substrate being 1cm×1cm (the availa-ble area for resonator is already smaller than this, as it should be placed somewhere inside the pick-up loop); for instance, at typical feature size of 100 μm, a complementary split double ring resonator needs a diameter at least about 12 mm to resonate below 1 GHz with respect to the one proposed in [14], while we are using an area much smaller than this. This helps the resonator to have a stronger coupling to the SQUID, as the coupling coefficient $k$ between the SQUID and the resonator is highly improved when the equivalent radius of the resonator becomes equal to that of the SQUID [21]. Using small feature sizes in a spiral resonator helps extending the number of turns with the same equivalent size. It also works for arbitrary shapes resembling the one shown in Figure 3(a) to extend their length. This resonator is designed to have a maximized length in the area between the input loop and the pick-up loop. The pattern includes a long transmission line interrupted by some pseudo-circles to better couple to the SQUID.

Several configurations are explored under two major categories (resonator inside or outside the input loop), and the pattern with the highest quality factor and strongest coupling to the SQUID is selected for fabrication due to the requirement $k^2Q > 1$ [21]. The Q factor is calculated based on S11 (reflection to the port connected to the readout) and the coupling is calculated based on S21 (between the readout port and a second port placed instead of the Josephson junction). Several types of readouts including inductive ones such as square-shape loops in different sizes and circular loops with different radiuses, as well as transmission lines similar to the one shown in Figure 6 are examined to feed each resonator, and only results from the optimum one with the highest quality factor and coupling efficiency are reported for each resonator here.

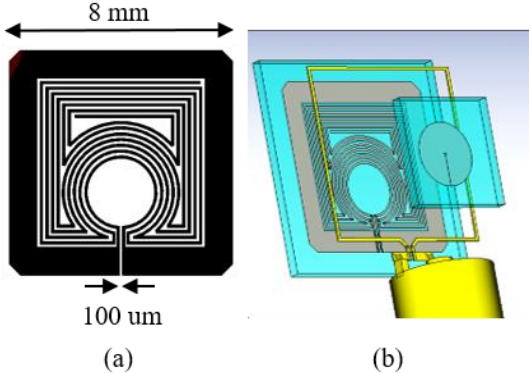

Figure 3. (a) Arbitrary shaped transmission line resonator between input loop and pick-up loop, (b) measurement configuration including readout loop and SQUID

### A. Resonator outside the input loop structure

In a monolayer structure, arbitrary patterns and split ring resonators (SRR) are capable of being placed in the area between the pick-up loop and the input loop, but not complementary split ring resonators (CSRR) or spiral shapes, because the flux transformer lines will interrupt them. SRR in the area between the input loop and the pick-up loop is studied thoroughly in [20]. The resonance frequency in this case (two rings and feature size of 100 μm) is around 1.5 GHz [20].

Even by extending the number of rings to 5 or using smaller feature size of 50 μm, the resonance frequency still does not go below 1 GHz.

Arbitrary patterns give the possibility to use the area between the input and the pick-up loops more efficiently. They are employed to achieve resonance frequencies lower than 1 GHz by extending the length of the strip-line and therefore increasing the inductance. One of these patterns is shown in Figure 3(a). The flip-chip configuration together with the readout loop, connected to a coaxial cable is also depicted in Figure 3(b). All of the aforementioned types of readouts are tested to feed this pattern and the optimum one for this case was a square-shape loop with a side of 5.3 mm. Figure 7 reveals the frequency response of this structure with a resonance frequency at 610 MHz and quality factor of 400. At this frequency, the value of S21 is around -40 dB, which is the lowest estimation of $k$. Actually, the value of $k$ between the resonator and the SQUID is higher than this value, as the S21 describes the coupling between the readout port and the junction, while some leakage is included due to the coupling between the readout and the resonator.

### B. Resonator inside the input loop structure

Either ring resonators or spiral shapes are potential fits for the resonator inside the input loop structure. A superconducting CSRR with 9 rings (almost filling the input loop area), and linewidth and feature size of 50 μm on a 1 mm thick LaAlO$_3$ substrate (Figure 4), and also a spiral resonator with almost the same filling factor (9.25 turns), and the same linewidth and feature on the same substrate (Figure 5) are fed by all mentioned types of the readouts.

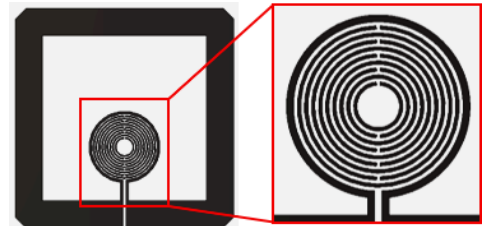

Figure 4. Complementary split ring resonator (CSRR) with 9 rings and feature size of 50 μm inside input loop

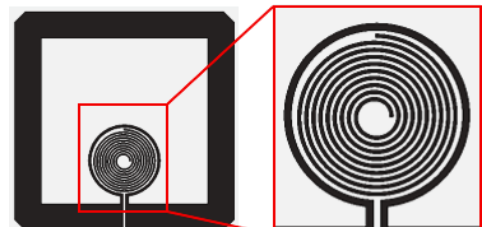

Figure 5. Spiral resonator with 9 turns and feature size of 50 μm inside input loop

The optimum readout for CSRR is obtained to be a circular coil with radius of 1 mm. According to Figure 8, the resonance frequency of the CSRR occurs at 2.347 GHz with a quality factor of 710. The resonance frequency is higher than that of the arbitrary pattern, because of the smaller length of this res-onator. Maximum S21 in this case reaches -7 dB, showing a jump about 33 dB with respect to the previous case. Two major reasons are responsible for this growth: first, the circular ring structure helps concentrating the magnetic field at the center, where the Josephson junction is located. Second, the arbitrary pattern is placed out of the input loop, therefore, the magnetic field that it induces at the Josephson junction should first pass over the input loop, which attenuates this field.

Figure 9 shows results from the spiral shape inside the input loop.

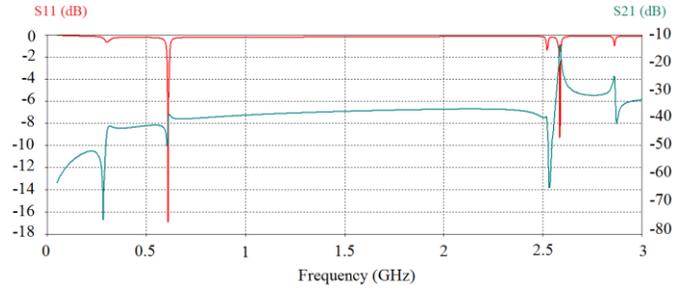

Figure 7. S parameters for Arbitrary shape resonator, S11 (red) ranging from -18 to 0 dB, S21 (blue) ranging from -80 to -10 dB

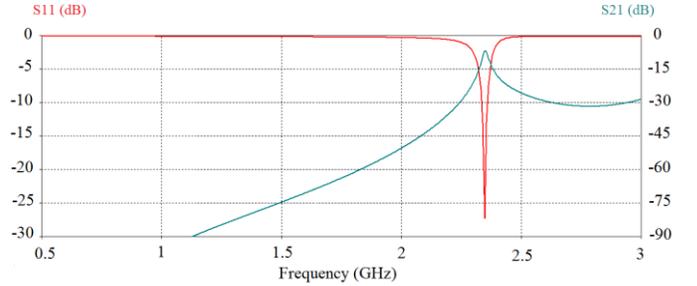

Figure 8. S parameters for SRR inside input loop, S11 (red) ranging from -30 to 0 dB, S21 (blue) ranging from -90 to 0

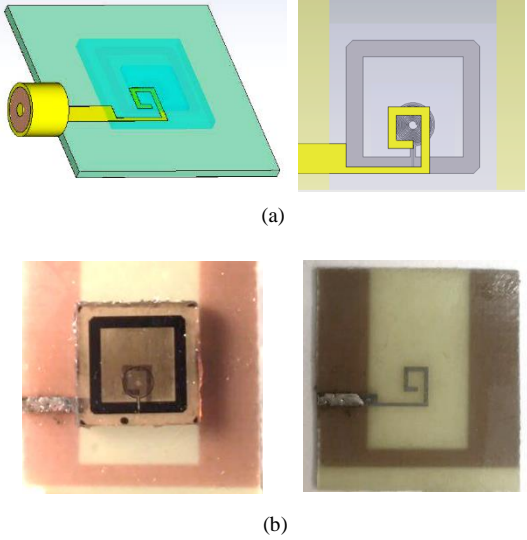

(a)

(b)

Figure 6. Transmission line (combined capacitive- inductive) readout configuration with the resonator substrate, (a) schematic, (b) fabricated

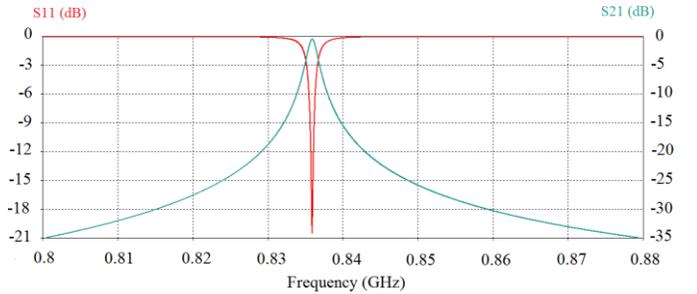

Figure 9. S parameters for spiral shape inside input loop, S11 (red) ranging from -21 to 0 dB, S21 (blue) ranging from -35 to 0 dB

The optimum readout for this resonator is the transmission line of Figure 6. Its resonance frequency is at 836 MHz with quality factor of 5900. This is much higher than previous Q factors introduced in [14] and [20] to the best of our knowledge. The coupling efficiency in this case is higher than all the previous ones, with a maximum S21 value of -0.5 dB. There is again some further enhancement, as the resonance frequency is now below 1 GHz despite the same size as the CSRR. The reason is that the effective length of the spiral is longer than the CSRR as the rings are not connected in the latter. Also, having such a low resonance frequency with these dimensions, means that the spiral is strongly coupled to the larger parts, which is the flux transformer being responsible for the low frequency behavior of the structure. The value of S21 is improved as well, which is again due to enhanced coupling of the spiral to the flux transformer as well as its capability of focusing the magnetic field at its center like a multi-turn coil. It should also be mentioned that all of the Q factors are reported by the loaded values, i.e., in presence of SQUID which suppresses the Q factor.

Finally, the spiral shape is fabricated using 200 nm tick YBCO film deposited on a crystalline 1 mm thick LaAlO$_3$ substrate. The frequency response of the fabricated sample is measured by the transmission line readout made by printed circuit board technology, and the results were in agreement with the simulations.

## IV. CONCLUSION

We proposed a monolayer optimum design of HTS flux concentrator integrated with resonator for strong coupling to rf-SQUIDs with about 3 millimeters diameter washer areas. An optimized monolayer flux concentrator was obtained with pick-up and input loops chosen so that to keep their inductances as close as possible to achieve the maximum matching and coupling efficiency possible. We investigated three different types of resonators on a 1cm×1cm area LaAlO$_3$ substrate and compared all of their specifications. This was to reach the optimal resonator pattern with highest quality factor and strongest coupling between the resonator and the SQUID with a resonance frequency below the maximum affordable frequency of the electronic system used. Various types of readouts were tested to find the optimum operation for each pattern. The spiral pattern shows a resonance frequency below 1 GHz featuring optimal values for the quality factor and the coupling coefficient compared to those of the others. The spiral shape had all of the optimal specifications together with having a resonance frequency at 836 MHz and the quality fac-tor of 5900 in addition to a maximum obtained coupling coefficient value of -0.5 dB.

ACKNOWLEDGMENT

The authors would like to thank Roya Mohajeri and Rana Nazifi for conducting the fabrication process, and Arman Alizadeh for the FEM simulation assistance.